\documentclass{llncs}
\usepackage[english]{babel}
\usepackage{graphicx}
\usepackage{amssymb, amsmath}

\title{Trees with small b-cromatic index}
\author{Ana Silva}
\institute{ParGO Group - Parallelism, Graphs and Optimization\\Universidade Federal do Cear\'a, Fortaleza, Brazil\\\email{anasilva@mat.ufc.br}}

\begin{document}

\maketitle

\begin{abstract}
In a recent article~\cite{JP.14}, the authors claim that the distance between the b-chromatic index of a tree and a known upper bound is at most 1. At the same time, in~\cite{MS.12} the authors claim to be able to construct a tree where this difference is bigger than 1. However, the given example was disconnected, i.e., actually consisted of a forest. Here, we slightly modify their construction in order to produce trees, thus getting that indeed the difference between the b-chromatic index of trees and the known upper bound can be arbitrarily large. We also point out the mistake made in~\cite{JP.14}.
\end{abstract}

Given a proper coloring $\psi:V(G)\to \{1,\cdots,k\}$ of $G$, we say that color $i\in \{1,\cdots,k\}$ is \emph{realized in $\psi$} if there exists $u\in V(G)$ such that $\psi(u) = i$ and $\psi(N[u]) = \{1,\cdots,k\}$; we also say that \emph{$u$ realizes color $i$ in $\psi$}, and that $u$ is a \emph{b-vertex of $\psi$}. A \emph{b-coloring} of the vertices of a graph is a proper coloring that realizes every color. Observe that if $\psi$ does not realize color $i$, then each vertex $u$ in color class $i$ can have its color modified to some $j\in \{1,\cdots,k\}\setminus \psi(N[u])$. This produces a proper color that uses fewer colors, which implies that any coloring with $\chi(G)$ colors must be a b-coloring. Therefore, we are actually interested in the worst case scenario of such a coloring, and define the \emph{b-chromatic number of $G$} as the maximum integer $b(G)$ for which $G$ has a b-coloring with $b(G)$ colors. This problem was introduced by Irving and Manlove in~\cite{IM.99}, where they also showed that computing $b(G)$ is $\mathcal{NP}$-hard in general and polynomial-time solvable for trees. The problem is still $\mathcal{NP}$-hard when restricted to bipartite graphs~\cite{KRATOCHVIL.etal.02}, chordal distance-hereditary graphs~\cite{HLS.11}, and line graphs~\cite{Campos.etal.15}.

 In their seminal paper, Irving and Manlove also introduced an important upper bound for this metric. Given a b-coloring with $k$ colors, observe that there must exist at least $k$ vertices in $G$ with degree at least $k-1$ (namely, the b-vertices). Therefore, if $m(G)$ is defined as being the largest integer $k$ such that $G$ has at least $k$ vertices with degree at least $k-1$, then if follows that: \[\chi(G)\le b(G)\le m(G)\]

They then proved that if $G$ is a tree, then $b(G)\ge m(G)-1$. Their proof also gives a polynomial algorithm that produces an optimal b-coloring of a tree. A natural question arises whether the same holds for trees on the edge-coloring version of the problem. In other words:

\begin{question}\label{q} Is it true that $b(G)\ge m(G)-1$, whenever $G$ is the line graph of a tree (or equivalently a claw-free block graph)? \end{question}
 
In~\cite{MS.12}, the authors say that the answer is ``no'' and claim to have a construction of line graphs of trees where the distance $m(G)-b(G)$ is arbitrarily large. However, their construction produced disconnected graphs, which clearly cannot be line graphs of trees. Later, in~\cite{JP.14}, the authors claim that the answer to the question is ``yes'', but their proof also had some problems. 

Here, we modify the construction presented in~\cite{MS.12} to produce a connected claw-free block graph. This gives that the answer to Question \ref{q} is indeed ``no''. In Section~\ref{error}, we point out what is the problem in the proof presented in~\cite{JP.14}. We mention that the edge version of the problem restricted to trees has recently been settled and that it is polynomial~\cite{CS.15}. 

We say that $u\in V(G)$ is \emph{dense} if $d(u)\ge m(G)-1$, and denote the set of dense vertices of $G$ by $D(G)$. A graph is called \emph{tight} if $\lvert D(G)\rvert = m(G)$ and $d(u) = m(G)-1$, for every $u\in D(G)$. Note that if $G$ is tight and $b(G) = m(G)$, then each vertex in $D(G)$ is a b-vertex.


\section{Graphs with small b-chromatic number}
\label{construction}

In this section, we show that the b-chromatic number of a connected claw-free block graph $G$ can be arbitrarily smaller than $m(G)$. 

\begin{figure}
\begin{center}
\includegraphics[height=6cm]{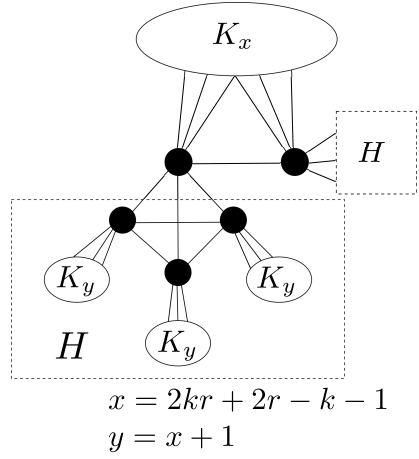} 
\end{center}
\caption{Representation of an $r,k$-gadget.}
\label{fig:construction}
\end{figure}

Let $r$ be any positive integer, and $k$ be an integer such that $k> r$. As you read the construction, you should  observe Figure \ref{fig:construction}. We also advise that you set $r=2$ and $k=3$ to get the smallest example $T$ that can be constructed here with $b(T) < m(T)-1$. An {\it $r,k$-gadget} is the graph $G'$ obtained from a clique $C$ of size $2kr+2r-k-2$, adding two adjacent vertices $v',v''$ that are also adjacent to every vertex of the clique, then adding two cliques $S',S''$ of size $k$, with $S'$ adjacent to $v'$ and $S''$ to $v''$, and, finally, for each vertex $u\in S'\cup S''$, adding a clique of size $2kr+2r-k-1$ adjacent to $u$. We denote the starting clique of $G'$ by $C(G')$, the two vertices adjacent to $C(G')$ by $v^1(G')$ and $v^2(G')$, and the clique adjacent to $v^j(G')$ by $S^j(G')$, $j=1,2$.

Let $G$ be the disjoint union of $r$ copies of the $r,k$-gadget, $G_1,\cdots,G_r$. Finally, we need to connect these copies. For each copy $G_i$, consider a vertex $u^1_i\in S^1(G_i)\setminus \{v^1(G_i)\}$ and a vertex $u^2_i\in S^2(G_i)\setminus \{v^2(G_i)\}$. Then, choose vertices $w^j_i\in N(u^j_i)\setminus (S^j(G_i)\cup\{v^j_i\})$, for each $i\in \{1,\cdots,r\}$ and $j\in\{1,2\}$. Add edges $(w^2_i,w^1_{i+1})$, for $i\in \{1,\cdots,r-1\}$. 

In what follows, we want to show that $G$ is tight, and that $b(G)\le m(G)-r$. For this we suppose that it is not the case and get that all the dense vertices in some of the gadgets, say $G_i$, must be b-vertices. We get a contradiction by counting the number of colors that must appear in $C(G_i)$ and comparing it with the number of vertices in the clique. We observe that some mofications in the construction can be done, as long as these steps still hold. For instance, if we do not need $G$ to be connected, we can ignore the last step of the construction, in which case we may consider $k = r$.  Now, we proceed to our proof. First, note that, for each $i\in\{1,\cdots,r\}$, we have: 

\begin{itemize}
\item a vertex $u\in C(G_i)$ has degree equal to $2kr+2r-k-2-1+2=2kr+2r-k-1$; 
\item $d(v^j(G_i))=2kr+2r-k-2+1+k=2kr+2r-1$, for $j=1,2$; 
\item for all $u\in S^j(G_i)$, $j=1,2$, $d(u)=k+2kr+2r-k-1=2kr+2r-1$; and 
\item $d(u)\le 2kr+2r-k-1+1 = 2kr+2r-k$, for every other vertex $u$. 
\end{itemize}

Observe that there are $r(2k+2)=2rk+2r$ vertices with degree $2kr+2r-1$, and every other vertex has degree at most $2kr+2r-k \le 2kr+2r-(r+1)$, which is strictely smaller since $r\ge 1$. Therefore, $m(G)=2kr+2r$ and the set of vertices with degree at least $m(G)-1$ is exactly $\bigcup_{i=1}^r(S^1(G_i)\cup S^2(G_i)\cup\{v^1(G_i),v^2(G_i)\})$; denote this set by $D$. Now, let $V'\subseteq V(G)$ be the basis of a b-coloring of $G$ with at least $m(G)-r+1$ colors. Since $d(u)\le 2kr+2r-k \leq m(G)-r-1$, for each $u\in V(G)\setminus D$, we get that $V'\subseteq D$. Also, since $|V'|>m(G)-r$, there must exist a gadget $G_i$ such that $D\cap V(G_i)\subseteq V'$. Note that, because $v^1(G_i)$ and $v^2(G_i)$ are b-vertices and $N(v^j(G_i))\setminus V'=C(G_i)$, $j=1,2$, the colours of all vertices in $\mathcal{V}'=V'\setminus \{v^1(G_i),v^2(G_i)\}$ must appear in $C(G_i)$. However, $|\mathcal{V}'|\ge m(G)-r-1$, while $|C(G_i)|=2kr+2r-k-2 \le m(G) - (r+1) - 2 = m(G)-r-3$, a contradiction.


\section{Error in \cite{JP.14}}\label{error}

The algorithm presented in~\cite{JP.14} is an adaptation of the algorithm for trees presented by Irving and Manlove in their seminal paper. We give an outline of the algorithm for trees and say where it fails when adapted to the edge version.

Consider a tree $T$. For any vertex $v\in V(T)$, denote by $dist_2(v)$ the set of vertices at distance exactly 2 from $v$. We say that $T$ is \emph{pivoted} if $\lvert D(T)\rvert = m(T)$ and there exists $v\in V(T)\setminus D(T)$ such that:
\begin{itemize}
  \item $D(T)\subseteq N(v)\cup dist_2(v)$; and
  \item If $u\in dist_2(v)$, then there exists $w\in N(u)\cap N(v)$ such that $d(w) = m(T)-1$.
\end{itemize}

One can verify that if $T$ is a pivoted tree, then $b(T) < m(T)$. This is because, supposing otherwise, we end up being forced to repeat a color in $N(u)$, for some $u\in D(T)$ with $d(u) = m(T)-1$. In~\cite{IM.99}, they show how to color a pivoted tree with $m(T)-1$ colors. After this, they show that if $T$ is not pivoted, then $T$ has a convenient subset of dense vertices which can play the role of the b-vertices in a b-coloring of $T$ with $m(T)$ colors. This convenient subset is called a good set, and it is defined as follows.

Given a subset $W\subseteq D(T)$ and a vertex $v\in V(T)\setminus W$, we say that $W$ \emph{encircles $v$} if $W\subseteq N(v)\cup dist_2(v)$ and, for every $u\in dist_2(v)\cap W$, there exists $w\in N(v)\cap N(u)\cap W$ with $d(w) = m(T)-1$. We call $W$ a \emph{good set} if $\lvert D(W)\rvert = m(T)$, $W$ does not encircle any vertex, and $w\in N(W)$, for every $w\in V(T)\setminus W$ such that $d(w) \ge m(T)$.

As we said before, Irving and Manlove prove that if $T$ is not pivoted, then $T$ has a good set $W$; then, they show how to construct a b-coloring with $m(T)$ colors having $W$ as b-vertices. Let $T'= T[N[W]]$. First, note that if $\psi$ is a b-coloring of $T'$ with $m(T)$ colors, then a b-coloring of $T$ can be greedily obtained since $d(w)<m(T)$, for every $w\in V(T)\setminus N[W]$. Therefore, the difficulty in coloring $T$ lies in the coloring of $T'$. Irving and Manlove also noticed that only some of the vertices in $T'$ are hard to color; these, they called inner vertices. A vertex $x\in N(W)\setminus W$ is called \emph{inner} if $x$ is within a path of length at most 3 between two vertices of $W$; if $x\in N(W)\setminus W$ is not an inner vertex, then it is called an \emph{outer} vertex.

The algorithm in~\cite{IM.99} starts by coloring each $w\in W$ with a distinct color. Then, they color the inner vertices ensuring that not too many colors are repeated in $N(w)$, for every $w\in W$. Finally, they argument that the outer vertices in $N(w)$ can be freely colored with the colors missing in $N(w)$. Because not too many colors are repeated during the coloring of the inner vertices, it is possible to prove that each $w\in W$ can be turned into a b-vertex in this phase.

Now, translating this to edges, Jakovac and Peterin defined an edge $e\in U = N(W)\setminus W$ as being ``\emph{inner} if there exists two edges $e_i,e_j\in W$, at distance at most 2 from each other, with $e$ on the path between them''. Then, they proceed to use the same strategy as the one used in~\cite{IM.99}. However, they did not notice that an outer edge in this case may be adjacent to more than one edge in $W$ (in our construction, these would be the edges that form the cliques $C(G_i)$ of each gadget $G_i$). This means that, if $uv$ is an outer edge and $E' = N(uv) \cap W = \{uv_1,\cdots,uv_q\}$, it cannot be ensured that there will be a color that is missing in the neighborhood of each $uv_i\in E'$ with which we can color $e$. Therefore, it is not true that at the end of the procedure each edge in $W$ realizes a distinct color.

We mention that, in her thesis~\cite{S.10}, Silva had already noticed that good sets are not so good for block graphs. There, considering $G$ to be a block graph and $W$ to be a good set in $G$, if $u\in N(W)\setminus W$ is not an inner vertex and has more than one neighbor in $W$, then $u$ is called a \emph{side vertex}. It is proved that if $W$ is a good set in a block graph $G$ and $W$ has no side vertices, then $W$ can play the role of b-vertices in a b-coloring of $G$. She also gives other sufficient conditions under which a good set produces a b-coloring. 

It is now known that, even if $b(T)$ is not always at least $m(T)-1$, when $T$ is the line graph of a tree, we can still  decide in polynomial time whether $b(T)\ge k$ for given $k$~\cite{CS.15}. The algorithm however does not give a structural characterization of $T$. Therefore, one can still ask what is the structure of a tree $T$ that has b-chromatic index exactly $m(T)$ (or at most $m(T)-k$, for some $k$).

\end{document}